# Improved IIR Low-Pass Smoothers and Differentiators with Tunable Delay

Hugh L. Kennedy
Defence and Systems Institute (DASI), School of Engineering
University of South Australia
Adelaide, Australia
hugh.kennedy@unisa.edu.au

*Abstract*—Regression analysis using orthogonal polynomials in the time domain is used to derive closed-form expressions for causal and non-causal filters with an infinite impulse response (IIR) and a maximally-flat magnitude and delay response. The phase response of the resulting low-order smoothers and differentiators, with low-pass characteristics, may be tuned to yield the desired delay in the pass band or for zero gain at the Nyquist frequency. The filter response is improved when the shape of the exponential weighting function is modified and discrete associated Laguerre polynomials are used in the analysis. As an illustrative example, the derivative filters are used to generate an optical-flow field and to detect moving ground targets, in real video data collected from an airborne platform with an electro-optic sensor.

*Keywords—Digital filters, IIR filters, Low-pass filters, Moving target indication, Recursive estimation*

I. INTRODUCTION

The fact that classical regression analysis, using orthogonal polynomials, automatically satisfies a number of highly desirable maximally-flat constraints means that digital smoothers and differentiators may be concurrently designed and visualized in the complementary time and frequency (or $z$-plane) domains [1]-[3]. Much of the early work on the engineering applications of such filters focused on the time domain [4], whereas the more recent signal-processing literature places a greater emphasis on the frequency domain [5],[6].

Low-pass maximally-flat designs are appealing because the specified response is guaranteed at the design frequencies; furthermore, closed-form expressions can usually be derived for the filter coefficients. Early maximally-flat finite-impulse-response (FIR) designs were formulated using flatness constraints with unity and zero gain at $\omega = 0$ and $\omega = \pi$, respectively [7], although the latter constraint may be relaxed for non-linear phase designs with a tunable group delay [8]-[10]. Infinite-impulse-response (IIR) designs have received somewhat less attention [11]. Closed form expressions for IIR filter coefficients that satisfy flatness constraints are given in [12]; however the form of the filter response is difficult to predict at non-design frequencies.

The causal IIR filters, derived in the time domain using discounted least-squares, in [4] were intended for use in tracking radar systems. Predictive forms with a low-frequency phase lead for a negative group delay are favored in these applications to allow the antenna control system to keep the target within the field of regard. The "fading memories" of such filters place the greatest weight on the most recent sample. It is shown in this paper that the frequency response may be improved, in systems where a moderate delay is tolerable, if the exponential weight used in the (recursive) regression analysis is replaced by a more general weight, which is closer to being symmetric and has a maximum at a non-zero delay. Similar weighting functions have been used to improve the frequency response (i.e. narrower main-lobe and lower side-lobes) of recursive analyzers used in wide-band frequency analysis [13]-[15].

Closed-form expressions for the filter coefficients are derived and presented in Section II; some tuning considerations are discussed and the filter responses are analyzed in Section III. The main result of this paper, as illustrated in Section III and summarized in Section IV, is a demonstration of the improved high-frequency noise-attenuation that is achievable, in cases where a larger group delay is tolerable, when *associated* Laguerre polynomials are used (the $\kappa = 1$ case), relative to traditional fading-memory smoother and differentiator designs (the $\kappa = 0$ case) [4], [16]. Non-causal extensions of the traditional approach are also provided for completeness. As an illustrative example, the derivative filters are used in Section V to generate an optical-flow field and to automatically detect moving ground targets, in real video data collected from an airborne platform with an electro-optic sensor.

## II. METHOD & RESULTS

Derivation of linear-difference-equation (LDE) coefficients using regression analysis in the time and/or space domains, with polynomial or sinusoidal models, has previously been used to design low-order IIR filters [17]. The complete process will therefore not be repeated here; however, a broad overview is given in this Section, to introduce key concepts, design parameters, and filter characteristics. The procedure is extended here to include derivative filters and non-zero shape parameters ($\kappa \geq 0$).

A continuous-time input $y(t)$, is sampled (i.e. measured) at time instants $nT$, where $T$ is the sampling period and $n$ is the sample index. Over a specified 'time-scale' in the vicinity of $n$, the following model is used to represent the signal structure and the measurement process:

$$x(n-m) = \sum_{k=0}^{B} \beta_k(n) \psi_k(m) \quad (1a)$$

$$y(n) = x(n) + \varepsilon \quad (1b)$$

where: $y(n)$ is the $n$th 'noise-corrupted' measurement; $x(n)$ is the corresponding 'noise-free' signal at time $nT$; $\varepsilon$ is a Gaussian-distributed noise term, with $\varepsilon \sim \mathcal{N}(0, \sigma_\varepsilon^2)$; $\beta$ are the local model coefficients; $B$ is the model degree; $\psi_k(n)$ is the $k$th local basis function at time $nT$; and $m$ is a delay index. The *discrete* basis functions are constructed by orthonormalizing a set of polynomial components, using a linear combination

$$\psi_k(m) = \sum_{i=0}^{B} \alpha_{k,i} m^i \quad (2)$$

where, in the general case, the $\alpha$ coefficients are determined using the Gram-Schmidt procedure such that

$$\sum_{m=0}^{\infty} \psi_{k_2}(m) w_+(m) \psi_{k_1}(m) = \delta_{k_1 k_2} \quad (3a)$$

in the causal case and

$$\sum_{m=-\infty}^{+\infty} \psi_{k_2}(m) w_\pm(m) \psi_{k_1}(m) = \delta_{k_1 k_2} \quad (3b)$$

in the non-causal case, where $\delta$ is the Kronecker delta function and $w(m)$ is a (non-normalized) weighting function with

$$w_+(m) = m^\kappa e^{\sigma m} \quad (4a)$$

in the causal case and

$$w_\pm(m) = e^{\sigma |m|} \quad (4b)$$

in the non-causal case; in both cases, $\sigma < 0$ for stable filter realizations.

The mean of (4a), for continuous $m$, is at $m = -(\kappa + 1)/\sigma$, thus older samples receive greater emphasis in the analysis as the "forgetting factor" $\sigma$, approaches zero (from the left) and as the shape parameter $\kappa$, increases. Orthonormalization, yields the discrete Laguerre polynomials for $\kappa = 0$ [4]; the discrete *associated* Laguerre polynomials result for $\kappa \geq 0$. In the non-causal case, the centroid of the weighting function is at zero for all parameter combinations, due to the use of a two-sided weighting function.

The model coefficient vector (or the "Laguerre spectrum" [4]) is determined via discounted least-squares *analysis*, using

$$\hat{\beta}_k(n) = \sum_{m=0}^{\infty} \psi_k(m) w_+(m) y(n-m) \quad (5a)$$

in the causal case and

$$\hat{\beta}_k(n) = \sum_{m=-\infty}^{+\infty} \psi_k(m) w_\pm(m) y(n-m) \quad (5b)$$

in the non-causal case. The maximum-likelihood estimate of the $D$th derivative of the input sequence is $\hat{x}_D(n)$. It is evaluated at time $T[n-q]$, using the model parameters $\hat{\beta}_k(n)$ in the *synthesis* equation:

$$\hat{x}_D(n-q) = \left(\frac{-1}{T}\right)^D \sum_{k=0}^{B} \hat{\beta}_k(n) \frac{d^D}{dm^D} \psi_k(m)|_{m=q}. \quad (6)$$

Note that the "hat" accent is used here to denote an estimated quantity. Fortunately, operations (5) & (6) may be combined and applied recursively by taking $\mathcal{Z}$ transforms. The discrete-time transfer function of the resulting causal filters, linking the $\mathcal{Z}$ transform of the input measurements $Y(z)$, to the $\mathcal{Z}$ transform of the output estimates $\hat{X}(z)$, i.e. $H(z) = \hat{X}(z)/Y(z)$, is

$$H(z) = \sum_{m=0}^{M_b-1} b_m z^{-1} / \sum_{m=0}^{M_a-1} a_m z^{-1}. \quad (7)$$

The transfer function has repeated (real) poles at $z = p$, where $p = e^\sigma$, and a pole multiplicity of $B + \kappa + 1$; the $q$ parameter only influences the zero locations. Following the process described in this Section yields the causal and non-causal filter coefficients given in Tables I-III. The coefficients in Table I may also be determined using (13.3.11) in [4]; the smoother (for $q = 0$) is also given in [18]. Improvements to these designs are made in this paper by generalizing to $\kappa \geq 0$ (see Table III). The non-causal smoothers and differentiators considered in [19] are also generalized here to higher-order cases (see Table II). Non-causal filters are realized by summing the outputs of two filters that are independently applied in the forward (FWD, increasing $n$), and backward (BWD, decreasing $n$) directions. The way in which the design parameters ($B, D, \kappa, p$ & $q$) affect the response of the filters is discussed in Section III, although many of the filter characteristics may be understood using the constructs of (discounted) regression analysis used in this Section.

TABLE I.  LDE COEFFICIENTS (CAUSAL, $B = 2, \kappa = 0$)

|   | Smoother [a] | Differentiator [b] |
|---|---|---|
| $c$ | $\frac{1}{2}(1-p)$ | $\frac{1}{2T}(1-p)^2$ |
| $b_0$ | $c(q^2p^2 + 3qp^2 + 2p^2$ $-2q^2p + 2p$ $+q^2 - 3q + 2)$ | $c(2qp + 3p$ $-2q + 3)$ |
| $b_1$ | $-c(2q^2p^2 + 8qp^2 + 6p^2$ $-4q^2p - 4qp + 6p$ $+2q^2 - 4q)$ | $-4c(qp + 2p$ $-q + 1)$ |
| $b_2$ | $c(q^2p^2 + 5qp^2 + 6p^2$ $-2q^2p - 4qp$ $+q^2 - q)$ | $c(2qp + 5p$ $-2q + 1)$ |
| $b_3$ | 0 | 0 |
| $a$ | $[1, \quad -3p, \quad 3p^2, \quad -p^3]$ | |

To place a zero at $z = -1$, for maximum high-frequency attenuation use:

[a.] $q = \left[4p - \sqrt{2(p^2 + 4p + 1)} + 2\right]/[2(1 - p)]$

[b.] $q = (1 + 2p)/(1 - p)$.

TABLE II.  LDE COEFFICIENTS (NON-CAUSAL, $B = 2, \kappa = 0$)

|   | Smoother (FWD&BWD) | Differentiator (FWD/BWD) |
|---|---|---|
| $c$ | $1/[2(p^2 + 8p + 1)]$ | — |
| $b_0$ | $c(p^2 + 10p + 1)(1-p)/(1+p)$ | 0 |
| $b_1$ | $3cp(p^2 - 1)$ | $(+/-)(p-1)^3/[2T(p+1)]$ |
| $b_2$ | $3cp^2(p^2 - 1)$ | 0 |
| $b_3$ | $cp^3(p^2 + 10p + 1)(1-p)/(1+p)$ | 0 |
| $a$ | $[1, \quad -3p, \quad 3p^2, \quad -p^3]$ | $[1, \quad -2p, \quad p^2, \quad 0]$ |

TABLE III. LDE Coefficients (Causal, $B = 2, \kappa = 1$)

| | Smoother [c] | Differentiator [d] |
|---|---|---|
| $c$ | $\frac{1}{6}(1-p)^2$ | $\frac{1}{2T}(1-p)^3$ |
| $b_0$ | 0 | 0 |
| $b_1$ | $c(3q^2p^2 + 9qp^2 + 6p^2$ $-6q^2p + 6qp + 12p$ $+3q^2 - 15q + 18)$ | $c(2qp + 3p$ $-2q + 5)$ |
| $b_2$ | $-2c\left\{\begin{matrix}(qp+3p)\\(-q+3)\end{matrix}\right\}\left\{\begin{matrix}(3qp+3p)\\(-3q+3)\end{matrix}\right\}$ | $-4c(qp + 2p$ $-q + 2)$ |
| $b_3$ | $c(3q^2p^2 + 15qp^2 + 18p^2$ $-6q^2p - 6qp + 12p$ $+3q^2 - 9q + 6)$ | $c(2qp + 5p$ $-2q + 3)$ |
| $b_4$ | 0 | 0 |
| $a$ | $[1, \quad -4p, \quad 6p^2, \quad -4p^3, \quad p^4]$ | |

To place a zero at $z = -1$, for maximum high-frequency attenuation use:

[c.] $q = [4p - \sqrt{2(p^2 + 6p + 1)} + 4]/[2(1-p)]$;

[d.] $q = 2(1 + p)/(1 - p)$.

## III. Discussion

The resulting causal and non-casual filters have appealing frequency responses that approximately satisfy various constraints (maximally flat). The validity of the approximations improve as $\omega = 0$ is approached, where $\omega$ is the angular frequency $\omega = 2\pi f$ (radians per sample) and $f$ is the normalized frequency (cycles per sample). For a smoothing filter ($D = 0$) the frequency response is flat, with unity magnitude and linear phase (for a group delay of $q$ samples). For a differentiating filter ($D = 1$) the frequency response has linear magnitude $|H(\omega)| = \omega$ and phase. For both filter types, the frequency range, over which these frequency-domain properties are approximately true, increases with $B$.

The frequency response $H(\omega)$, of the filter is found by substituting $z = e^{j\omega}$ into (7). Using the causal smoother filter coefficients given in Table I in (7) and evaluating derivatives of $|H(\omega)|^2$ at $\omega = 0$, reveals that the first, second and third derivatives are all equal to zero, confirming that the procedure does indeed result in some degree of flatness.

The gradual roll-off of these maximally-flat filters makes it difficult to clearly specify and identify pass-band, transition-band and stop-band regions, which are central to the more conventional equi-ripple and weighted-integral-squared-error (WISE) design processes used in FIR filter design [20],[21]. If the polynomial basis set represents the low-frequency content of the signal, with all other frequency components due to noise $\varepsilon$, then the $p$ (or $\sigma$) parameter determines the ability of the filter to discriminate between the two sub-spaces, i.e. the attenuation at 'far-from-zero' frequencies. Noise power in the filter output decreases as more data are considered in the analysis process (i.e. as $\sigma \to 0$ for $p \to 1$). This improves frequency selectivity but decreases temporal selectivity due to a lengthening of the impulse response, which is not ideal for handling input discontinuities. IIR filters are particularly efficient in 'very-low-pass' roles because the time scale of analysis does not affect the order of the LDE.

The $q$ parameter adjusts the gain and phase characteristics of the causal filters ($q = 0$ for all non-casual filters). In some applications, closed-loop control systems for instance, the ability to manipulate the group delay at low-frequencies is critical ($q > 0$ for a phase lag or $q < 0$ for a phase lead); in other applications, audio processing for example, it is more important to strongly attenuate high frequencies. The proposed filters may be constructed in one of two ways, depending on design priorities: Either the $q$ parameter is arbitrarily chosen to yield the desired delay (see Fig. 1 and Fig. 3). (Note that in all phase-response sub-plots, lines of constant group delay, equal to $q$ samples, are plotted to give an indication of phase linearity.) Alternatively, an 'optimal' $q$ value is determined for a given $p$, using the footnotes to Tables I and III, to place a zero at $z = -1$ for infinite (dB) loss at $\omega = \pi$ (see Fig. 2 and Fig. 4) which also minimizes the variance reduction factor [17]. In the former design case, a reasonable value of $q$ must be chosen to ensure that there is sufficient data 'support', or analysis weight afforded by $w_+(m)$, before *and* after the synthesis point at $n - q$ to promote the desirable qualities discussed so far in this Section. This phenomenon is well known in regression analysis, where estimation/prediction errors are modelled using Student's t distribution or Snedecor's F distribution for uniformly weighted data over a finite interval [22]. It is surprising that these classical relationships are not utilized in recent studies on the time-domain properties of Savitzky-Golay smoothers and differentiators [23],[24].

For $\kappa = 0$, i.e. pure exponential decay, the value of $q$ directly determines the number of samples that follow the synthesis sample in the analysis; however if $q$ is increased too far, the relative weight applied in the vicinity of the synthesis point is diminished, which *degrades* the frequency selectivity of the filter (see Fig. 1 and Fig. 2). This effect is most pronounced for near-zero $p$ where the weight decays rapidly. It also explains why predictive filters (with $q < 0$) amplify high-frequency noise [25]. As a guide, $q$ should be kept near the centroid of $w_+(m)$. Using $\kappa > 0$ provides the opportunity to use an analysis weight that is more symmetric around a delayed synthesis point, which *improves* the frequency selectivity of the filter (compare Fig. 3 and Fig. 4 with Fig. 1 and Fig. 2).

The responses of some alternative FIR and IIR smoothers, are provided for comparison in Fig. 5. These methods were selected: firstly, because they place an emphasis on low-frequency flatness; and secondly, because they enable the filter coefficients to be found without iterative optimization procedures.

The minimized WISE design used a unity magnitude and a 1000x weight in the pass-band ($|f| \leq f_c$) to promote phase linearity, with zero gain and unity weight elsewhere; no transition-band was used. In time-domain target-tracking applications, the length of the finite memory filter is chosen to balance the contributions of random and systematic errors [4]. One of the interesting, and possibly undesirable, properties of the WISE FIR smoothers in Fig. 5 is the enhanced attenuation of sinusoids with periods that match the length ($MT$) of the analysis window, yielding a non-monotonic response. This is also a characteristic of finite-memory polynomial filters, e.g. FIR Savitzky-Golay filters [23]-[25]. The FIR-WISE filters are not maximally flat at $\omega = 0$; however, the flatness improves (at the expense of stopband attenuation) as the relative passband weight increases. The proposed smoothers only allow the phase and magnitude near $\omega = 0$ to be manipulated; however the WISE design method is ideal for arbitrarily-defined frequency bands.

The "universal maximally flat" (UMF) low-pass filters [10], with an FIR were also investigated and a linear-phase instantiation is also shown in Fig. 5. Placing all zeros at $z = -1$ gave a passband width that was similar to that of the corresponding linear-phase FIR-WISE filter. This design procedure provides exceptionally good high-frequency attenuation however the monotonic gain response makes it impossible to create a sharp transition between the passband and stopband.

The "maxflat fractional delay" (MFFD) filters described in [12] have excellent frequency responses with good magnitude flatness and phase linearity over a wide frequency range. This is achieved using low-order filters because poles may be placed arbitrarily in the complex $z$-plane. However not all combinations of numerator and denominator polynomial orders are able to satisfy the flatness constraints for a specified group delay, therefore a lengthy trial-and-error search is required to find an appropriate combination of parameters that yields a filter with the desired response (see Fig. 5).

Differentiator responses are plotted in Fig. 6. Note the following: the desired magnitude linearity in the low-frequency region for all filters; the increased attenuation at mid frequencies when $\kappa$ is increased from 0 to 1 (at the expense of a longer delay); the identical magnitude responses of the causal IIR filter (for $\kappa = 1$) and the non-causal IIR filter; and the reasonable phase linearity for all filters at low frequencies. Note also the improved noise attenuation for the IIR filters with optimal $q$ assignment at medium to high frequencies, relative to the FIR design [26]. Like the smoothing filters, some of this attenuation is sacrificed if $q$ is instead chosen to yield the desired group delay. It is difficult to attenuate mid-range frequencies for the maximally-flat FIR differentiator (and smoother) – adding more zeros at -1 offers diminishing returns. The IIR differentiators in [27] have very similar responses to the maximally flat FIR response shown and both are more suitable when a wide-band differentiator is required. IIR differentiator (and smoother) designs involving iterative optimization procedures, such as those described in [6], were not considered for comparison here.

Any number of alternative design techniques could have been used to design the types of low-pass IIR filters considered here, possibly resulting in superior properties with respect to a given design requirement – e.g. pass-band gain flatness, pass-band phase linearity, pass-band width, transition bandwidth and stop-band attenuation. However, by appealing to the concepts of discounted least-squares regression in the time domain, the main advantage of the proposed design approach is the ease with which: 1) closed-form expressions for the filter coefficients may be derived, at least in low-order cases; and 2) impulse and frequency (i.e. phase and magnitude) responses may be intuitively adjusted using two principal design parameters ($p$ and $q$) to achieve the desired effect. Thus complications arising from slow/non-convergent optimization procedures or continuous-to-discrete transformations, and even the use of computing aids, are avoided.

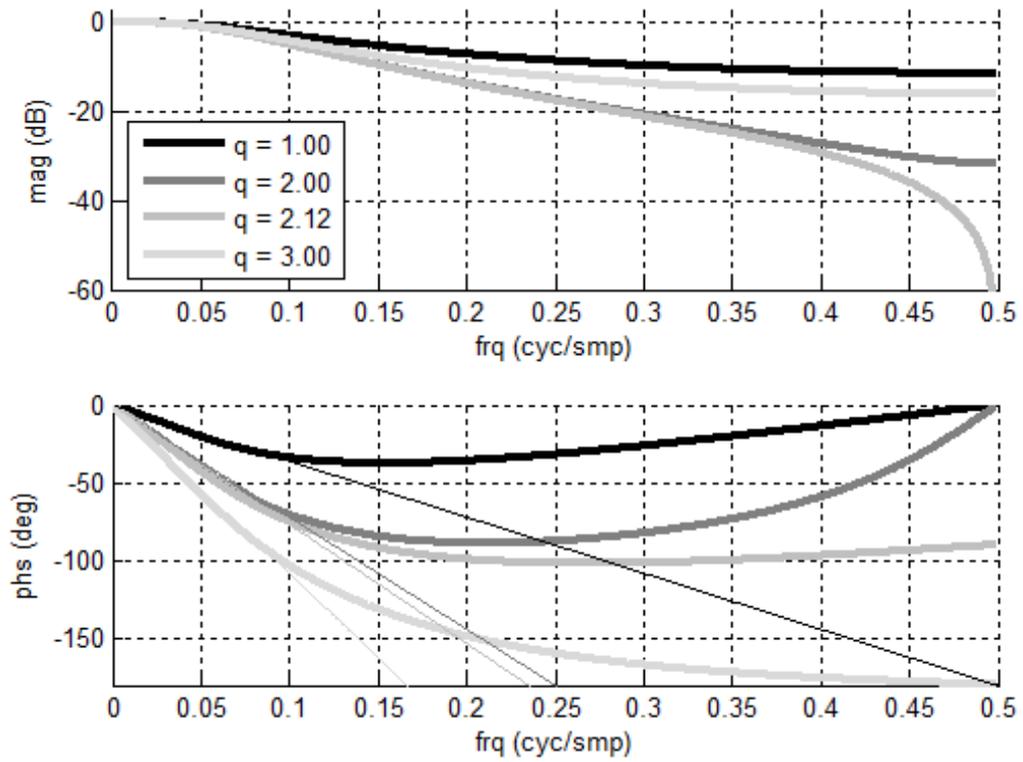

Fig. 1. Frequency response of causal IIR smoothers with $B = 2$, $\kappa = 0$ and $\sigma = -1/2$, as a function of $q$. Optimal response for $q = 2.12$.

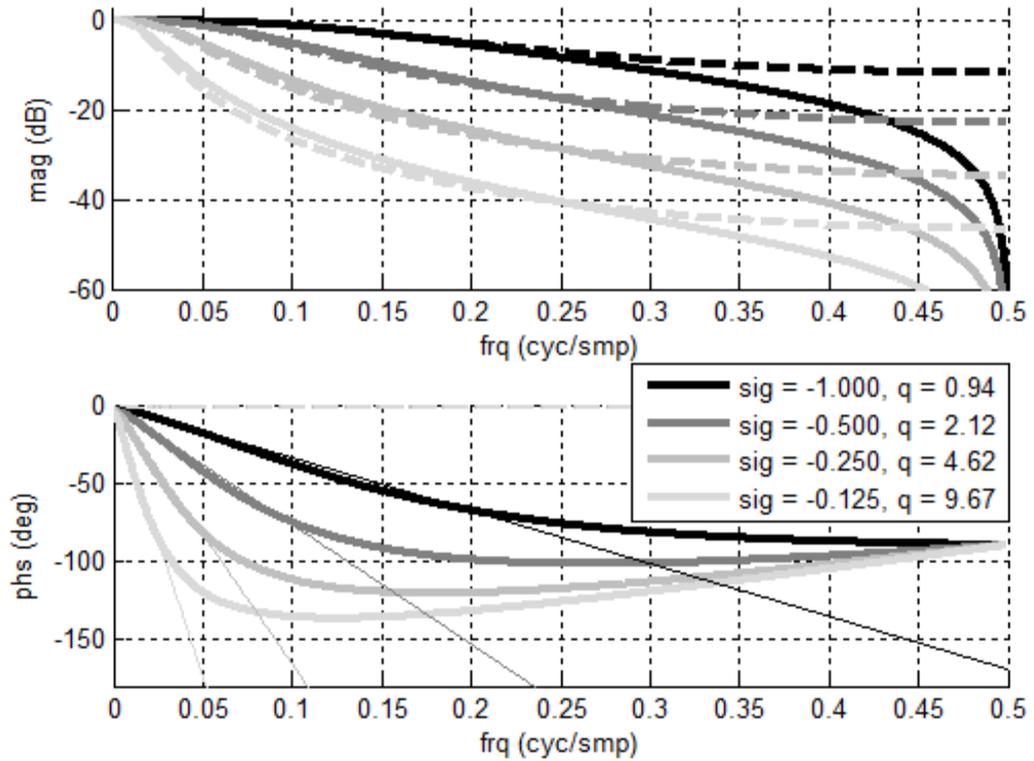

Fig. 2. Frequency response of IIR smoothers with $= 2$, $\kappa = 0$. Causal filters (solid lines) for a variety of optimal $\sigma$ and $q$ combinations; non-causal filters (dashed lines) for $q = 0$.

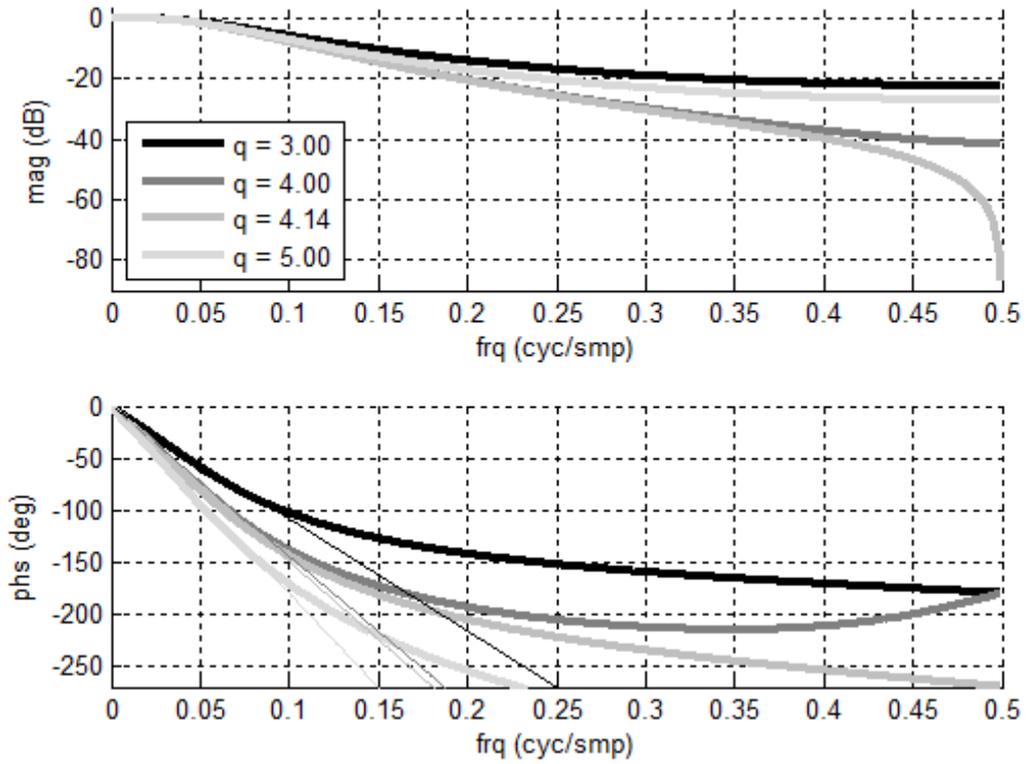

Fig. 3. Frequency response of causal IIR smoothers with $B = 2$, $\kappa = 1$ and $\sigma = -1/2$, as a function of $q$. Optimal response for $q = 4.14$.

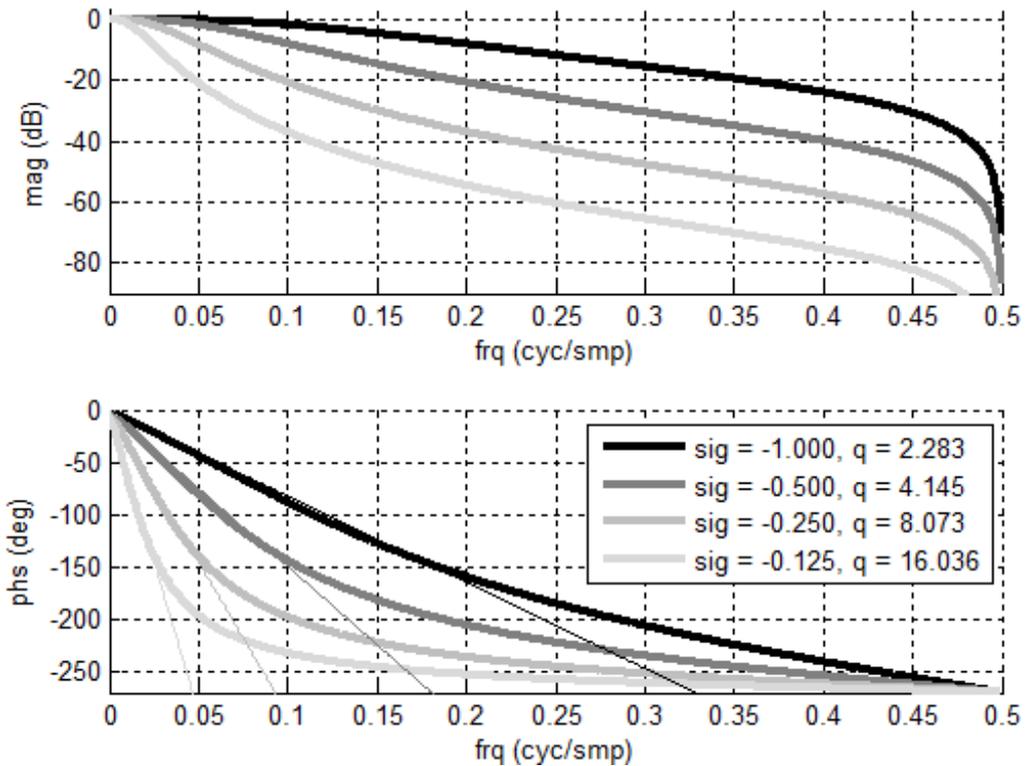

Fig. 4. Frequency response of causal IIR smoothers with $B = 2$, $\kappa = 1$ for a variety of optimal $\sigma$ and $q$ combinations.

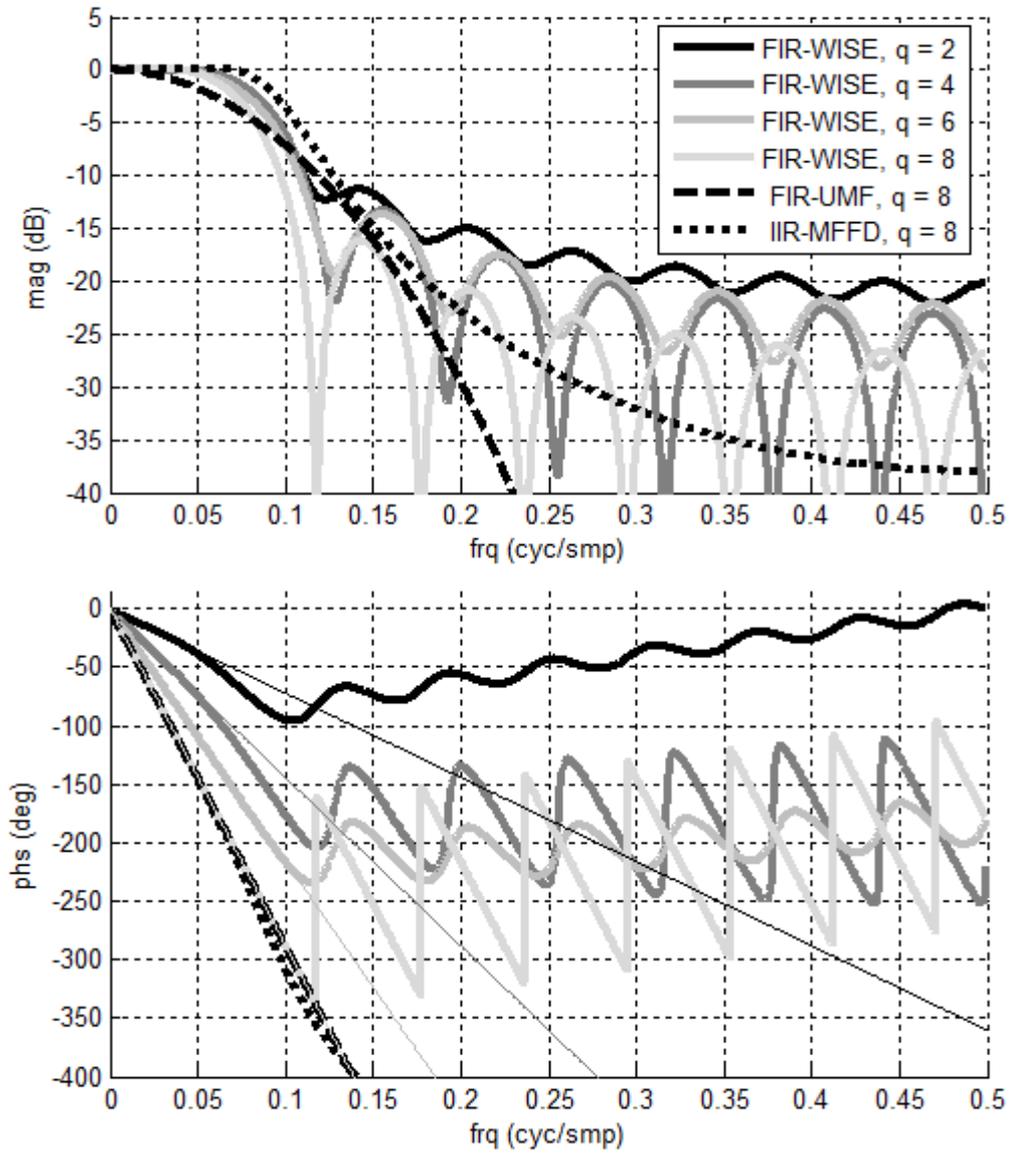

Fig. 5. Frequency response of various FIR and IIR smoothers. All FIR filters designed using $M = 17$. FIR-WISE filter designed using $f_c = 0.05$ and a 1000x relative passband weight, for various pass-band group delays, including the linear-phase case ($q = 8$). Linear-phase FIR-UMF filter designed with all zeros at $z = -1$. IIR-MFFD filter designed using $q = 8$, a numerator order of 1 and a denominator order of 4.

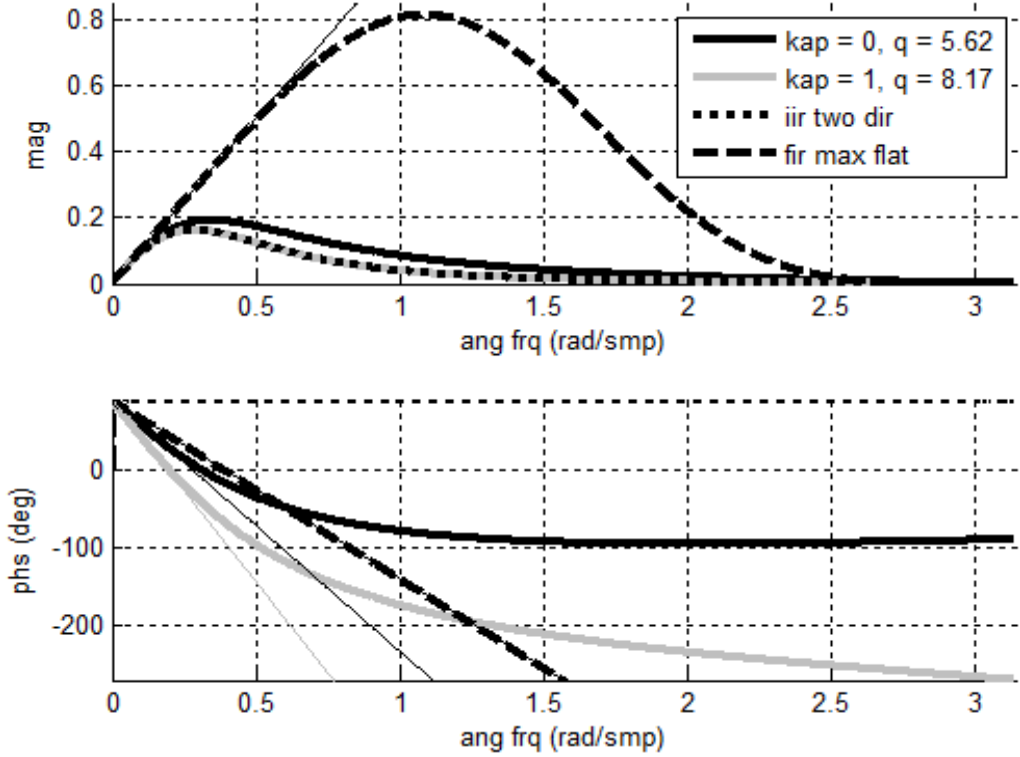

Fig. 6. Differentiator magnitude response (linear scale) and phase response. Maximally-flat linear-phase FIR filter (dashed line) with $M = 9$ and five zeros at $z = -1$; IIR filters designed using $B = 2$ and $\sigma = -1/2$. Causal variants (solid lines) with $\kappa = 0$ and $\kappa = 1$ (kap) and optimal $q$; non-casual variant (dotted line) also shown.

## IV. APPLICATION

The recursive IIR derivative filters described above were used in an algorithm to automatically highlight moving ground vehicles observed from an airborne platform with an electro-optic video camera. The raw data were collected at a rate of 25 frames per second, with red, green and blue channels, quantized using 24 bits at a resolution of 1920 by 1080 pixels per frame. One thousand frames of a 480 by 640 region-of-interest were then extracted and converted to a monochrome intensity map $I$, which was then (post) processed using a MATALB ® script, coded using only the core MATALB (R2013b) engine (i.e. no toolboxes), running on a personal computer with an Intel ® i7-4810MQ central processing unit with four cores (for eight concurrent threads) and a 2.8 GHz clock. A throughput rate of approximately 4 frames per second was achieved.

The optical-flow field of the (apparently) rotating and translating background was generated using the Lucas Kanade algorithm [28]-[31]. Spatial partial derivatives $I_x$ and $I_y$ were generated using *non-causal* filters with $B = 2$, $\kappa = 0$ and $q = 0$ (see Table II for filter coefficients); temporal partial derivatives $I_z$ were generated using *causal* filters with $B = 2$, $\kappa = 1$ and $q = 4$ (see Table III for filter coefficients). Both filter types used $\sigma = -1$ for $p = 0.3679$. Note that sample indices $n_x$, $n_y$ and $n_z$ have been omitted in this section for brevity; also note that a delay of 4 frames is applied to the spatial partial derivatives so that $I_x$ and $I_y$ are aligned with the output of the temporal filter, $I_z$. Local averages $J_{xx}, J_{xy}, J_{xz}, J_{yy}$, and $J_{yz}$, of the intensity partial-derivative products $I_x I_x$, $I_x I_y$, $I_x I_z$, $I_y I_y$, and $I_y I_z$, were then computed using non-causal and causal recursive exponential smoothers (i.e. $B = 0$ and $\kappa = 0$) with a single pole at $z = \exp(-1/16)$. The optical flow field of the background was then generated in the usual way using

$$\begin{bmatrix} v_x \\ v_y \end{bmatrix} = - \begin{bmatrix} J_{xx} & J_{xy} \\ J_{xy} & J_{yy} \end{bmatrix}^{-1} \begin{bmatrix} J_{xz} \\ J_{yz} \end{bmatrix} \quad (8)$$

where the velocity components $v_x$ and $v_y$ are expressed in pixels per frame units when $T = 1$ pixel or frame for the spatial and temporal filters, respectively.

The standard Lucas Kanade algorithm was extended in this work so that it is able to provide a means of detecting small moving targets that are set against a non-uniformly moving background. It can be seen from (8) that

$$\begin{bmatrix} J_{xz} \\ J_{yz} \end{bmatrix} = -\begin{bmatrix} J_{xx} & J_{xy} \\ J_{xy} & J_{yy} \end{bmatrix} \begin{bmatrix} v_x \\ v_y \end{bmatrix} \qquad (9)$$

represents the *averaged* spatiotemporal partial-derivative products due to the assumed background motion, thus

$$\begin{bmatrix} \mathcal{J}_{xz} \\ \mathcal{J}_{yz} \end{bmatrix} = -\begin{bmatrix} I_x I_x & I_x I_y \\ I_x I_y & I_y I_y \end{bmatrix} \begin{bmatrix} v_x \\ v_y \end{bmatrix} \qquad (10)$$

may be interpreted as being the local contribution of background motion to the *raw* spatiotemporal partial-derivative products. It therefore follows that

$$\Delta \mathcal{J} = \sqrt{(I_x I_z - \mathcal{J}_{xz})^2 + (I_y I_z - \mathcal{J}_{yz})^2} \qquad (11)$$

may be used as a convenient indication of the 'surplus' spatiotemporal partial-derivative products, due to local foreground motion, which cannot be accounted for by delocalized background motion.

An illustrative example of this processing architecture is provided in Figs 7-9. In this particular application, sensor noise power is very low; therefore, temporal filters designed using $\kappa = 0$ and $\kappa = 1$ produced similar optical flow fields and disparity/salience maps. Furthermore, the spatial and temporal filters are required to cope with discontinuities in polynomial model parameters $\beta$, due to object edges (e.g. buildings) therefore filters with a short impulse response were required, using poles close to the center of the unit circle ($p = 0.3679$).

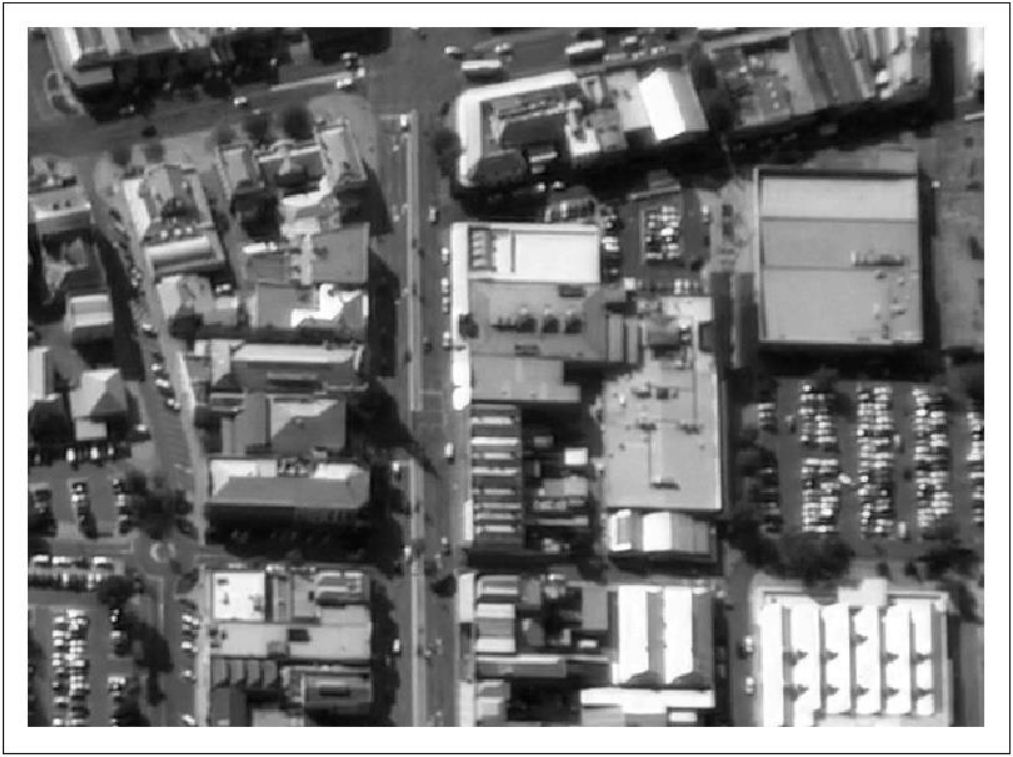

Fig. 7. A single frame of the monochrome region of interest $I$, containing an urban traffic scene, with moving vehicles and parked vehicles on the roadside.

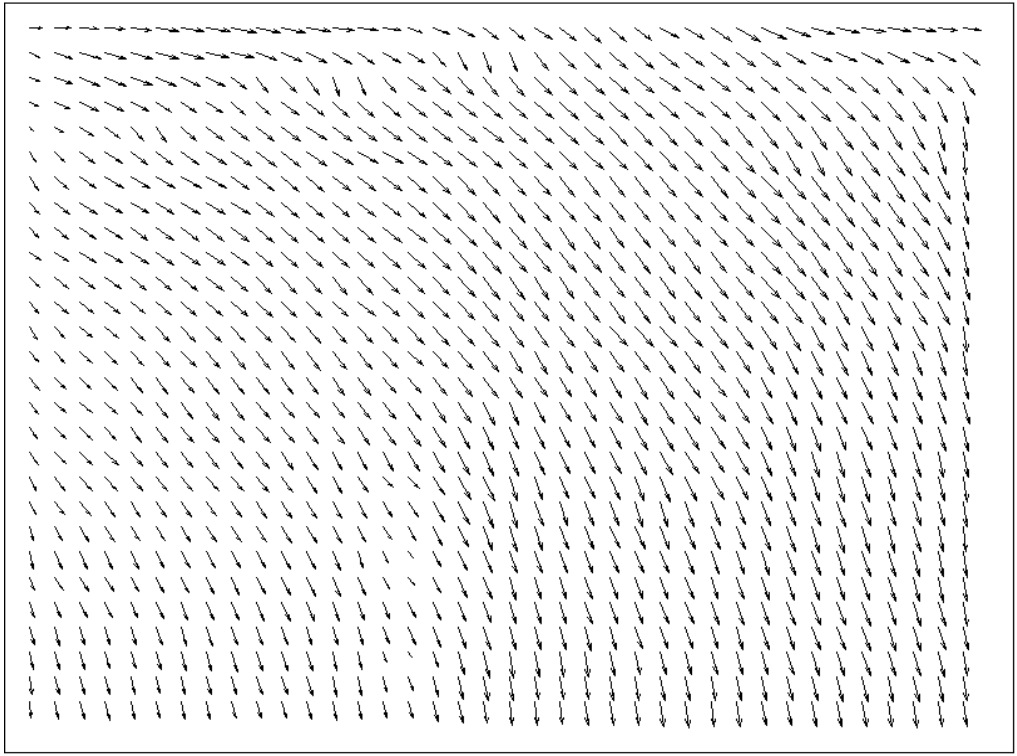

Fig. 8. Optical flow field ($v_x, v_y$) of the background. As the observing aircraft manoeuvres above the scene, the background appears to translate and rotate (around the bottom left-hand corner of the image). Away from the image edges (where evidence of spatial filter start-up transients are apparent) the velocity estimates appear to be reasonable.

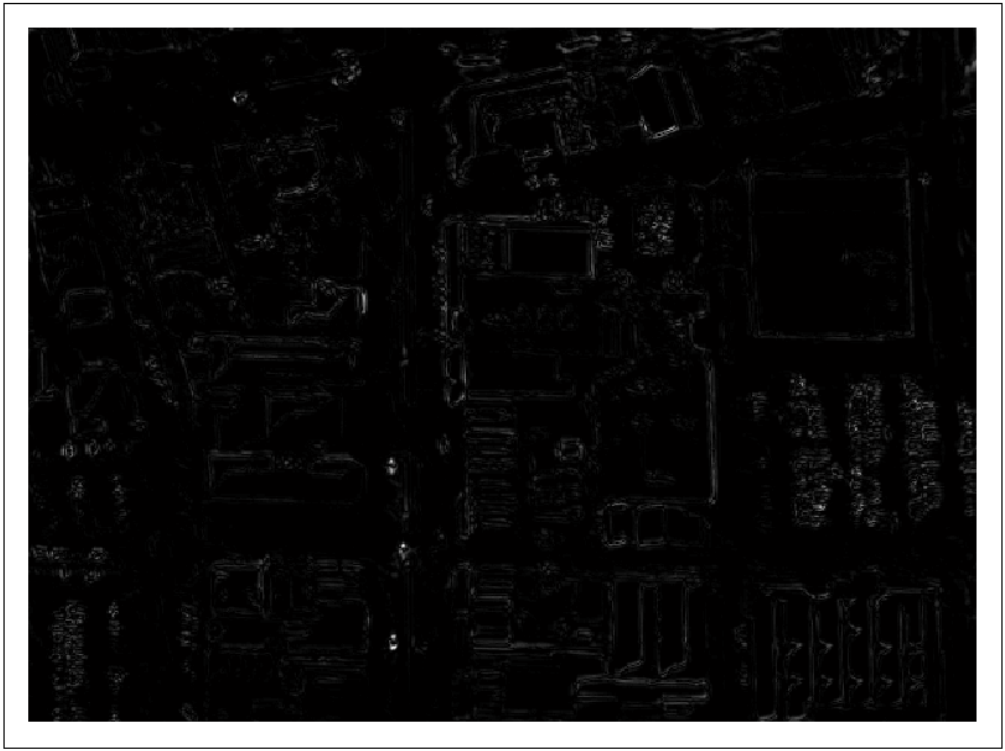

Fig. 9. Map of disparity/salience, $\Delta \mathcal{J}$. The three vehicles in the 'foreground' that are moving towards the top of the image, i.e. contrary to the motion of the 'background', are the brightest objects in the image.

## V. Conclusion

In this paper it is qualitatively shown that time-domain regression-analysis satisfies a number of highly desirable design constraints in the $\omega$ domain. This perspective naturally leads to the introduction of the shape parameter ($\kappa$), which allows more favorable phase/gain compromises to be reached. The IIR smoothing and differentiating filters presented here may find application in image-processing or machine-vision areas, more specifically, in systems that require:

- low-order filters for a high rate of data throughput, low-pass characteristics for the removal of high-frequency noise (e.g. as an alternative to simple frame differencing or for use in gradient-based optical flow calculation and/or moving target indication),
- a tunable impulse response duration (using $p$) to accommodate the tradeoff between steady-state frequency selectivity and transient response in non-stationary environments, and
- a tunable phase response (using $q$) to attain the desired balance between frequency selectivity and group delay.

The intended application motivated the consideration of both causal and non-causal filters.